\documentclass[sigconf,nonacm]{acmart}

\AtBeginDocument{%
  \providecommand\BibTeX{{%
    \normalfont B\kern-0.5em{\scshape i\kern-0.25em b}\kern-0.8em\TeX}}}

\copyrightyear{2021}
\acmYear{2021}
\setcopyright{acmcopyright}
\acmConference[SIGMOD '21]{Proceedings of the 2021 International Conference on Management of Data}{June 20--25, 2021}{Virtual Event, China}
\acmBooktitle{Proceedings of the 2021 International Conference on Management of Data (SIGMOD '21), June 20--25, 2021, Virtual Event, China}
\acmPrice{15.00}
\acmDOI{10.1145/3448016.3457545}
\acmISBN{978-1-4503-8343-1/21/06}

\usepackage{todonotes}
\usepackage{nicefrac}

\newcommand{\alt}[2]{
\addtolength{\jot}{-1ex} 
\begin{minipage}{#1} 
\vspace{4pt}
\begin{center}
$\begin{aligned}
#2
\end{aligned}$
\end{center}
\end{minipage}
}

\usetikzlibrary{arrows,positioning,backgrounds} 
\tikzset{
    rt/.style={
		rectangle,
		fill = white,
		draw=black, 
		text centered,
		inner sep=0.5ex
		},
	rtr/.style={
		rectangle,
		fill = white,
		draw=red, 
		text centered,
		inner sep=0.5ex
		},
	rtb/.style={
		rectangle,
		fill = white,
		draw=blue, 
		text centered,
		inner sep=0.5ex
		},
    rtt/.style={ 
    	rt,
    	inner sep=0.1ex
    	},
    ert/.style={ 
     	rt,
     	dashed
     	}, 
    ertr/.style={ 
     	rtr,
     	dashed
     	},  	
    ertt/.style={ 
        rtt,
        dashed
        }, 
    rect/.style={ 
        rectangle,
        fill = white,
        rounded corners,
        draw=black, 
        text centered,
        inner sep=0.8ex
        },
    rectw/.style={
        rect,
        draw=white
        },
    erect/.style={ 
    	rect,
    	dashed
    	},
    erectw/.style={ 
     	rectw,
     	dashed
     	},
    arrout/.style={
           ->,
           -latex,
           },
    arrin/.style={
           <-,
           latex-,
           },
    arrb/.style={
           <->,
           >=latex,
           }
}

\newcommand{\const}{\text{\rm {\bf Const}}}
\newcommand{\cL}{\mathcal{L}}
\newcommand{\cM}{\mathcal{M}}
\newcommand{\cG}{\mathcal{G}}
\newcommand{\cP}{\mathcal{P}}
\newcommand{\cV}{\mathcal{V}}
\newcommand{\lb}[1]{{\small \texttt{#1}}}
\newcommand{\lbe}[1]{{\small \texttt{#1\,=\,}}}
\newcommand{\bc}{\text{\it bc}}

\newcommand{\cat}{\text{\it concat}}
\newcommand{\ep}{\text{\it end}}
\newcommand{\stp}{\text{\it start}}
\newcommand{\test}{\text{\it test}}

\newcommand{\spanl}{\text{\sc SPanl}}
\newcommand{\p}{\text{\sc P}}
\newcommand{\np}{\text{\sc NP}}

\newcommand{\cA}{\mathcal{A}}
\newcommand{\cB}{\mathcal{B}}

\newcommand{\sem}[1]{\llbracket #1 \rrbracket_{\cL}}
\newcommand{\semp}[2]{\llbracket #1 \rrbracket_{#2}}

\newcommand{\ccp}{\text{\sc Count}}
\newcommand{\gen}{\text{\sc Gen}}

\newcommand{\FO}{\text{\rm FO}}
\newcommand{\FOC}{\text{\rm FOC}}

\newcommand{\COMB}{\text{\rm COMB}}
\newcommand{\AGG}{\text{\rm AGG}}
\newcommand{\CSL}{\text{\rm CSL}}

\allowdisplaybreaks

\begin{document}

\fancyhead{}

\title{Querying in the Age of Graph Databases and Knowledge Graphs}

\author{Marcelo Arenas}
\email{marenas@ing.puc.cl}
\affiliation{%
  \institution{Universidad Cat\'olica \& IMFD}
  \country{Chile}
}

\author{Claudio Gutierrez}
\email{cgutierr@dcc.uchile.cl}
\affiliation{%
  \institution{DCC, Universidad de Chile \& IMFD}
  \country{Chile}}

\author{Juan F. Sequeda}
\email{juan@data.world}
\affiliation{%
  \institution{data.world}
  \country{USA}
}


\begin{abstract}
Graphs have become the best way we know of representing knowledge. The
computing community has investigated and developed the support for
managing graphs by means of digital technology. Graph databases and
knowledge graphs surface as the most successful solutions to this
program. The goal of this document is to provide a conceptual map of the data
management tasks underlying these developments, paying particular
attention to data models and query languages for~graphs.
\end{abstract}





\maketitle

\section{Introduction}
\label{sec:why_tutorial}

What does it mean to query a graph?
What does it mean querying graph models? 
Any answers to these questions should try to understand the role of
graphs as a conceptual tool to model data, information and 
knowledge. 
Graphs have a long tradition as medium of representation, 
and an impressive wide range of uses.
Let us recall some highlights related to our area:
underlying data structures (the hierarchical and networks
database systems of the sixties \cite{stonebraker2005goes});
semantic networks; graph neural networks; entity relationship
model; XML; graph databases; the Web as universal network of information (and later of
data and knowledge); and knowledge graphs.
This non-exhaustive list indicates at least that some reflection is 
necessary before addressing the goal of this document.

\begin{figure}[h]
\includegraphics[width=8cm]{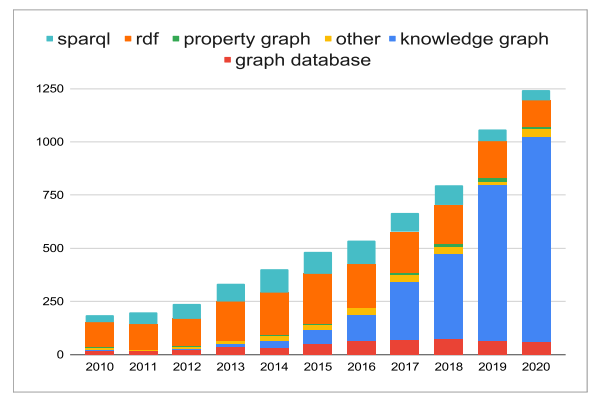}
\caption{Number of knowledge graph related publications. Source: DBLP.}
\label{fig:KGDBLP}
\end{figure}

Following usual practices, 
we performed an initial examination of what is being researched 
disciplinary in the area of data, graphs and knowledge.
We explored five of the most salient keywords that today 
represent research around this area:  ``graph database''; ``RDF''; ``SPARQL''; ``property graph'', ``knowledge graph'', by analyzing papers in computer science (those
indexed by DBLP)\footnote{This includes all types of publications indexed by DBLP.}
having these strings in their titles.\footnote{Data: https://data.world/juansequeda/dblp-knowledge-graph}
Figure~\ref{fig:KGDBLP} shows the evolution of the number of publications of papers with these keyword in the title from 2010 to 2020.

In this preliminary exploration we can observe the following.
The growth of ``knowledge graph'' papers can be seen starting in 2013, which correlates with the year after the Google's Knowledge Graph announcement. 
The amount of publications about ``RDF'' and ``SPARQL'' continue to be stable. However we observe a decline when compared to ``knowledge graph''. 
In 2015, 70\% of knowledge graphs papers were about RDF/SPARQL, while that went down to 14\% in 2020.
Papers about ``graph database'' are comparatively small and there is no significant growth, while papers about ``property graph'' are negligible.
The main takeaway message from this seems to be that publications about knowledge graphs are significantly increasing and in some
sense ``dominate'' the area. Thus, when addressing graphs as a model for data to knowledge, we cannot ignore the obvious knowledge graph hype.

This poses the question: What are knowledge graphs and what is 
their relation to graph databases? In addressing this question we should be
cautions about two extremes: On one hand, as Jeffrey Ullman wrote, avoid to get
``engaged in hand-wringing over the idea that we [the database discipline] 
are becoming irrelevant'' \cite{Ullman20}.
On the other hand, try to understand if there is someting really new
under this new hype about knowledge graphs.
Our preliminary hypothesis --one that we will follow here-- 
is that this rather vague notion with no clear borders (see e.g. 
Appendix A in \cite{DBLP:journals/corr/abs-2003-02320}) encompasses
a great variety of methods and practices dealing with data, information and knowledge, orbiting around the  gravitation center of graph models.

Our intention in this brief overview is to try to convey a rough
cartography of what is means to query in this new scenario,
and instill in the audience some doubts and
reflections we have about the development of our area as a whole.
We feel that this reflection is more important than ever today,
when big data, deep learning and other trends of the beginning of
the 21st century have shaken computing as we know it.

\subsection{A conceptual map of this document}

The motivating questions for this work are why the rise of graph databases and knowledge graphs? and what are new techniques and challenges behind this boom? To try to give answers to these questions, we start by establishing in Section \ref{sec:what} a
general picture of the area in order to better organize the conceptual
map of the wide diversity of existing methods and techniques. 

Then we continue by focusing on providing in Section \ref{sec:data_models} a unified and simple view of the data models behind graph databases and knowledge graphs, and show some recently established results on querying such graphs in Section~\ref{sec:query}. In particular, we focus on this latter section on queries about the local structure of a graph, the connections in it, and its global structure. Obviously, these are just a few tasks among many other relevant ones for graph databases and knowledge graphs: integration of different sources, graph data management, visualization of results, development of visual query languages, completion and refinement of graphs, search and exploration, reasoning in the presence of ontologies, different aspects related to natural language processing, analytics and the fundamental role of humans in all these challenges.
The reader should bear in mind that in this document, these topics are not covered with the same level of detail as data models and querying. 


%



\section{Databases, Graphs and Knowledge Graphs}
\label{sec:what}

Before going to technicalities, we will present a  conceptual
overview of the main notions of the area and how they interplay. 
Our goal is to contribute to a necessary discussion of the
reasons why graphs have become so prominent in data and knowledge
management.

\subsection{A necessary digression}

Data, information and knowledge appear so often in the literature,
and with so many senses, that we consider necessary to 
state how we understand them.
These concepts predate computer science, and there is a wide and rich literature about them \cite{furner,floridi,machlup,burke}.

The notion of {\em data} has at least three main senses that play a role in our discussion.
(a) A minimal unit of distinction. This lead to the notion of bit, that
that can be
implemented through electromagnetic media, and 
has many properties that makes it
flexible enough for storing, transmitting and manipulating.
(b) A register of aspects of reality. 
For example, the output of an instrument in a scientific experiment that captures some facets of ``reality''. 
(c) A mark that leaves a phenomenon of the world. For example, the footprint (now in a stone) of a dinosaur in an archaeological site, or today's registers from a sensor or a camera.

These three senses have something in common: data without further processing does not make much sense for humans, but it works as support for meaning and knowledge. 
Data in the computer science sense is this raw material in the digital format, 
incarnated as bits.
It is important to mention then that the form how data is organized,
e.g. the format and model for it, 
has to do essentially with the interpretation of registers in the modes of (b) and (c).

{\em Knowledge}, as understood e.g. in ``knowledge representation'', ``knowledge base'' or ``knowledge graph'', is a notion coming from the tradition of (formal) 
reasoning, and encompasses both, objects that statically represent knowledge (books, maps,charts, theorems, scientific laws, etc.), 
and mechanisms to dynamically 
obtain, deduce, or 
infer new knowledge from known premises or inputs (deductive systems, 
reasoners, neural networks, etc.). In fact, 
it is developed in \cite{GS21} the idea that 
a great part of the developments in data and knowledge since the 50s can be thought of as a way of giving digital support to the representations (e.g. tables, graphs, etc.)
and mechanisms of traditional reasoning
(e.g. reasoners, expert systems, etc.).


Finally, {\em information} (which is not our concern here) speaks about semantics, meaning and interpretation for the subject that will use it.
A database as such does not mean anything. It needs an interpretative apparatus, which is usually the schema and, more generally, some metadata. 
A digital text is a sequence of symbols. Again, needs a context, an interpretation, to get information from it. A similar phenomena occurs with a sequence of pixels: means nothing if not interpreted by
a human eye or displayed by a visualization software. 



\subsection{Graph data and querying}

Data is the raw material of our area.
 Databases originated in the need to store, keep safe and private, organize 
 and operate 
 in a efficient, reliable and permanent form, big quantities of data in computers.
From these basic and essential tasks, it was developed what is probably 
the most important functionality of databases, namely querying, that boils down to have friendly, expressive, and efficient languages for defining, updating, transforming and extracting data. That is why query languages play such a relevant role in our discipline.  

One of  
the landmark advances in the field was the notion of 
``data independence''~\cite{10.1145/362384.362685}.
As computing is essentially about communicating
humans and machines, or better, human knowledge and machine operation,
for methodological and practical reasons one should separate 
the physical level (the one revolving around the machine and data) from the 
logical level (revolving around the way humans model reality and knowledge).
Our working hypothesis is that graphs are an appropriate way of representing,
both, data and knowledge. And this would be the reason why graphs
began to be so prominent in this field of data and knowledge management.




  Graphs have a simple data structure consisting of nodes and edges, which
has the 
 nice  operational properties of expressing relations, presenting data
 in a rather holistic way (neither ordered nor sequentially), and 
 last but not least, having a flexible structure that
 permits  growing  and shrinking (adding/deleting nodes and edges)  and integration
 (of different graphs) in a natural way.

Over this nice data structure, floats a similarly nice group of conceptual ideas.
First, entities, represented by nodes; second, connectivity, represented by edges and paths; and third, emergent (global) properties 
that the structure generates.
Then it comes as no surprise that languages for querying graphs deal with
extracting information about these three main features (we will
discuss them in detail in Section \ref{sec:query}):
\begin{enumerate}
    \item[(i)] {\bf local properties} (nodes and neighborhoods), where  pattern matching and its extensions play a key role, being usually approached with logical methods;
    \item[(ii)] {\bf connectivity}, where paths and more complicated structures need to be extracted, usually by means of a mixture of regular expressions and logical methods; and
    \item[(iii)] {\bf global properties}, where the entire structure of a graph is considered, and that  essentially need different methods and approaches than those of (i) and (ii). Such methods are usually put under the graph analytics umbrella.
\end{enumerate}


\begin{figure*}[ht!]
\begin{center}
\resizebox{\textwidth}{!}{
\begin{tikzpicture}
\node[rt] (n1) {$n_1$ : \lb{person}};

\node[rt, right=2.5cm of n1] (n2) {$n_2$ : \lb{person}};

\node[rt, below=1cm of n2] (n5) {$n_5$ : \lb{infected}};

\draw[arrb,bend left=10] (n1) to 
node[ert, above] (e1) {$e_1$ $:$ \lb{lives}}
(n2);

\node[rt, below=1cm of n1] (n3) {$n_3$ : \lb{bus}};

\draw[arrout,bend right=10] (n1) to 
node[ert, right] (e2) {$e_2$ $:$ \lb{rides}}
(n3);

\node[rt, below=1cm of n5] (n4) {$n_4$ : \lb{person}};

\draw[arrout, bend left=10] (n4) to 
node[ert, below] (e3) {$e_3$ $:$ \lb{rides}}
(n3);

\draw[arrb, bend left=10] (n1) to 
node[ert, above] (e4) {$e_4$ $:$ \lb{contact}}
(n5);

\node[rect, right=1cm of n2] (pn1) {
\alt{55pt}{ 
 \lbe{name} & \lb{Alice}\\
 \lbe{age} & \lb{30}}};

\node[rt] (lpn1) at (pn1.north) {$n_1$ : \lb{person}};

\node[rect, right=3.5cm of pn1] (pn2) {
\alt{55pt}{ 
 \lbe{name} & \lb{Eve}\\
 \lbe{age} & \lb{60}}};

\node[rt] (lpn2) at (pn2.north) {$n_2$ : \lb{person}};

\node[rect, below=1cm of pn2] (pn5) {
\alt{55pt}{ 
 \lbe{name} & \lb{Claire}\\
 \lbe{virus} & \lb{covid19}}};

\node[rt] (lpn5) at (pn5.north) {$n_5$ : \lb{infected}};

\draw[arrb, bend left=15] (pn1) to 
node[below, erect] (pe4) {
  \alt{55pt}{
\lbe{date} & \lb{3/4/21}}}
(lpn5);

\node[ert] (lpe4) at (pe4.north) {$e_4$ $:$ \lb{contact}};

\draw[arrb, bend left=10] (pn1) to 
node[above, erect] (pe1) {
  \alt{55pt}{
\lbe{zip} & \lb{10002}}}
(pn2);

\node[ert] (lpe1) at (pe1.north) {$e_1$ $:$ \lb{lives}};

\node[rt, below=2cm of pn1] (pn3) {$n_3$ : \lb{bus}};

\draw[arrout, bend right=10] (pn1) to 
node[right, erect] (pe2) {
  \alt{55pt}{
\lbe{date} & \lb{3/7/21}}}
(pn3);

\node[ert] (lpe2) at (pe2.north) {$e_2$ $:$ \lb{rides}};
    
\node[rect, below=1cm of pn5] (pn4) {
\alt{55pt}{ 
 \lbe{name} & \lb{Bob}}};

\node[rt] (lpn4) at (pn4.north) {$n_4$ : \lb{person}};

\draw[arrin, bend right=10] (pn3) to 
node[below, erect] (pe3) {
  \alt{55pt}{
\lbe{date} & \lb{3/7/21}}}
(pn4);

\node[ert] (lpe3) at (pe3.north) {$e_3$ $:$ \lb{rides}};

\node[rt, right=2.5cm of pn2] (vn1) {$n_1$};

\node[rectw, above left=0mm of vn1] (vvn1) {
{\scriptsize
$\begin{pmatrix}
  \lb{person}\\
  \lb{Alice}\\
  \lb{30}\\
  \bot\\
  \bot\\
  \bot  
\end{pmatrix}$}};

\node[rt, right=3.5cm of vn1] (vn2) {$n_2$};

\node[rectw, above right=0mm of vn2] (vvn2) {
{\scriptsize
$\begin{pmatrix}
  \lb{person}\\
  \lb{Eve}\\
  \lb{60}\\
  \bot\\
  \bot\\
  \bot
\end{pmatrix}$}};

\draw[arrb,bend left=10] (vn1) to 
node[ert, above] (ve1) {$e_1$}
(vn2);

\node[rectw, above=0mm of ve1] (vve1) {
{\scriptsize
$\begin{pmatrix}
  \lb{lives}\\
  \bot\\
  \bot\\
  \lb{10002}\\
  \bot\\
  \bot
\end{pmatrix}$}};

\node[rt, below=2cm of vn1] (vn3) {$n_3$};

\node[rectw, below left=0mm of vn3] (vvn3) {
{\scriptsize
$\begin{pmatrix}
  \lb{bus}\\
  \bot\\
  \bot\\
  \bot\\
    \bot\\
  \bot
\end{pmatrix}$}};

\draw[arrout,bend right=10] (vn1) to 
node[ert, left] (ve2) {$e_2$}
(vn3);

\node[rectw, left=0mm of ve2] (vve2) {
{\scriptsize
$\begin{pmatrix}
  \lb{rides}\\
  \bot\\
  \bot\\
  \bot\\
  \lb{3/7/21}\\
    \bot
\end{pmatrix}$}};

\node[rt, below=0.7cm of vn2] (vn5) {$n_5$};

\node[rectw, right=0mm of vn5] (vvn5) {
{\scriptsize
$\begin{pmatrix}
  \lb{infected}\\
  \lb{Claire}\\
  \bot\\
  \bot\\
  \bot\\
  \lb{covid19}
\end{pmatrix}$}};

\node[rt, below=1.1cm of vn5] (vn4) {$n_4$};

\node[rectw, below right=0mm of vn4] (vvn4) {
{\scriptsize
$\begin{pmatrix}
  \lb{person}\\
  \lb{Bob}\\
  \bot\\
  \bot\\
    \bot\\
  \bot
\end{pmatrix}$}};

\draw[arrout, bend left=10] (vn4) to 
node[ert, below] (ve3) {$e_3$}
(vn3);

\node[rectw, below=0mm of ve3] (vve3) {
{\scriptsize
$\begin{pmatrix}
  \lb{rides}\\
  \bot\\
  \bot\\
  \bot\\
  \lb{3/7/21}\\
    \bot
\end{pmatrix}$}};

\draw[arrb, bend left=17] (vn1) to 
node[ert, below] (ve4) {$e_4$}
(vn5);

\node[rectw, below=0mm of ve4] (vve4) {
{\scriptsize
$\begin{pmatrix}
  \lb{contact}\\
  \bot\\
  \bot\\
  \bot\\
  \lb{3/4/21}\\
    \bot
\end{pmatrix}$}};

\node[rectw, below=5.1cm of e1] (mlg) {(a) A labeled graph};

\node[rectw, below=5.2cm of pe1] (mpg) {(b) A property graph};

\node[rectw, below=5.05cm of ve1] (mvg) {(c) A vector-labeled graph that};
\node[below=0mm of mvg] (amvg) {};
\node[right=-18.1mm of amvg] (mvg1) {represents the property graph in (b)};
\end{tikzpicture}}
\end{center}
\vspace{-10pt}
\caption{Three graph data models.\label{fig-all-g}}
\end{figure*}
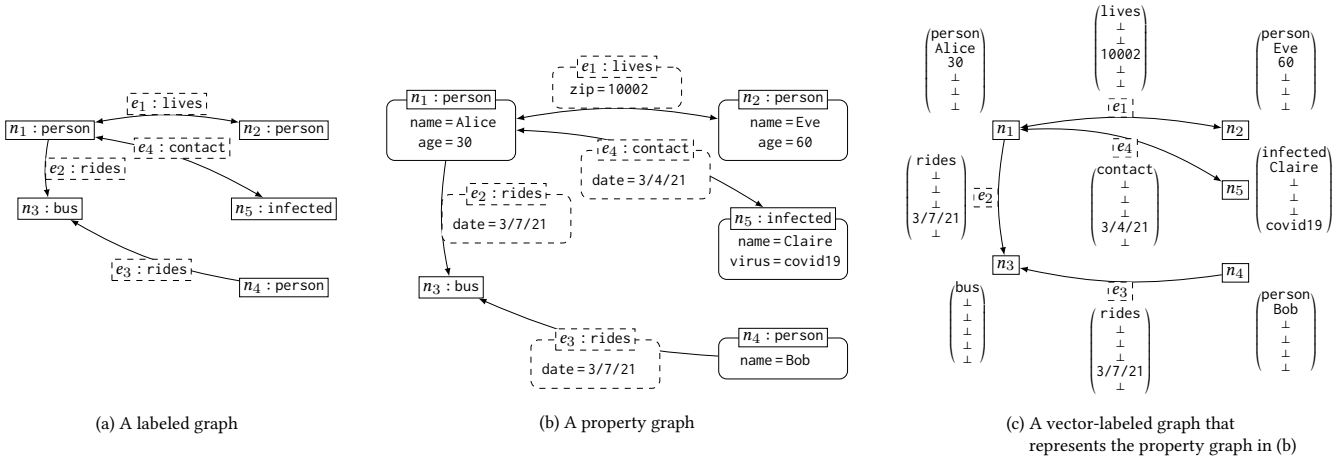

\subsection{Some paradigmatic examples}

To get a flavor of how these structures and ideas work in practice, let us 
briefly review three paradigmatic examples.

{\bf Semantic networks.}
 One of the first attempt to represent knowledge in the form of graphs was the notion of 
 Semantic Network~\cite{richens1956preprogramming,quillan1963notation,quillian1967word}. First born as a purely diagrammatic representation, soon researchers formalized those graphs to give them a formal  semantics using logical methods. Graphs (called networks in this field) are used because they are historically very good objects to represent knowledge (in the static sense). 
 Besides being a bridge to visualization of knowledge, 
 semantic networks have the feature (shared with graphs in general) of
 highlighting and facilitating the discovery and representation of relationships. 

{\bf Graph Databases.}
Classical relational databases are flexible enough to represent graphs,
e.g. using a two attribute relation representing edges in a graph. 
Moreover, one could consider 
relational schemata as generalization of binary relations ($m$-ary relations).
In this representation, nodes are entries and paths are constructed by successive
joins.
Why then do we need graph databases?
There are at least two reasons: joins are expensive and thus, reasoning about 
paths --that are one the main motive to use graphs-- becomes 
very costly.
But 
furthermore, global properties are not easy to compute with classical
queries based on logical methods.

So, it comes as no surprise that people have tried since the early days of
databases to develop graph databases.
First at the hardware level with the hierarchical and network
models of the sixties; then at a more logical level in the eighties
where we found the ``golden era'' of graph databases~\cite{ABDDMZ92,SWFKFUKCT93,10.1145/1322432.1322433}. However, many constraints, among them the limitations of hardware and software,
did not permit the popularization of these systems~\cite{10.1145/1322432.1322433}.

{\bf Semantic Web.}
The Web is the first comprehensive system for representing, integrating and ``producing'' knowledge at big scale, 
whose key idea is taking advantage of the structure and features of
the graph model (particularly the idea of integrating and extending 
knowledge and data by linking).

The Web was originally thought as a universal network, that is, nodes
 and edge were designed (with the notion of ``Web page'') 
 to be directly interpreted by humans. However, 
 due to its scale, soon emerged the need to enrich 
 the network to allow automation of functionalities. 
  This is the origin of the
 {\em Semantic Web} (SW), where semantics means ``understandable by machine''
 \cite{bernerslee2001semantic}.
 The SW contained the two ideas that were to originate the current notion of
 knowledge graph: first, a graph structure to organize the data; 
 second, a system which encodes not only information (i.e. data to be interpreted
 directly by humans), but it is also 
 organized to 
  derive new facts from the current ones, that is, deals 
essentially with knowledge.
 The feature of the underlying network devised by its creators, its universality  
 (given by URIs~\cite{berners1998uniform} and the hypertext transfer protocol HTTP), 
 became a  problem  for private companies
 and organizations as it was not ``practical'' due to privacy and property rights concern.
 Google overcame this, and develop the notion of knowledge graph as a ``finite'',
    manageable, controlled and usually private Semantic Web.
Nevertheless, the main concepts, namely, ontologies to integrate knowledge,
and the Web as medium, would become almost standards.

\subsection{Knowledge Graphs}

The cases of Semantic Networks, Graph databases and the Semantic Web
point to systems or software infrastructure in which graphs 
supports 
the representation, integration and production of knowledge.
This rather fuzzy idea is what is at the core of the notion of 
Knowledge Graph.
So we can define a Knowledge Graph as a software object (artifact) 
that represents (codifies), integrates and produces knowledge. In order to perform
these functionalities, it relies on the model of graphs.
In this regard, the notion of knowledge graph encompasses many previous developments \cite{FSAHKPTUW20,GS21}.

{\em Knowledge representation} is incorporated mainly through standardized languages,
metadata and ontologies on one hand, and semantic networks and other
non-formal language representations on the other. 
Lately, low-level representations like vector-labeled graphs were added. Notice that  all of them use the flexibility of graphs
for model representation (see Section \ref{sec:data_models}).

{\em Integration of knowledge} is achieved essentially via the aforementioned 
features of graph extensibility and integration (assuming a good and 
standardized representation like ontologies). Thus a knowledge graph is capable of
integrating knowledge from different sources, by linking 
or materializing them in one place, and in widely different formats, 
which include classical
tabular data and files, natural language text, images, sound, video 
and other types of sensor data.

Finally, {\em producing new knowledge} is probably the main ``added-value'' as 
opposed to classical repositories of knowledge.
Knowledge graphs have the capabilities of: 
 deducing, e.g. by means of logical reasoners or neural networks; linking, that is, relating different pieces of knowledge beforehand
isolated; learning, through new data and learning algorithms; and of course, generating new knowledge by human intervention through 
refined ways of querying it
(see section \ref{sec:query}). This new knowledge is not only user-oriented, but
is utilized in parallel to ``complete'' and enrich the knowledge already present.
As a clear example of this, we see the rapid development of knowledge graph embeddings~\cite{BUGWY13,CZC18}, and its use in the refinement and completion of knowledge graphs~\cite{LLSLZ15,P17}. In fact, they have even been used to enrich such knowledge by answering complex logical queries, which might need to deal with multiple unobserved nodes and edges in a graph \cite{HBZJL18,RHL20}.

These features make a knowledge graph a highly multifaceted software object.
One of the best examples of this is Wikipedia.
It is a repository that represents knowledge in the form of a large
graph (implemented via Web link protocols).
It organizes knowledge  like a classic encyclopedia, the paradigm of 
knowledge source in pre-computing era. But the greatest feature of
Wikipedia is its functionality of integrating knowledge 
from different sources (such as editors or links to other web pages) in different formats, 
both physical (text, images, sound, video, etc.)  
and in terms of semantics (different alphabets, languages, etc.). This makes it an
unbounded networks of knowledge. Moreover, when coming to generation of knowledge,
Wikipedia has a mechanism to ``extract'' or ``produce''
knowledge. In this case, these are the multiple interfaces, most of them 
human interfaces. Wikipedia is a knowledge graph oriented toward
a final human user, that is, it does not (or at least was 
not conceived to) feed another software systems. DBpedia~\cite{ABKLCI07} and Wikidata~\cite{VK14}
are derived system oriented to supply this facet.


\section{Graph data models}
\label{sec:data_models}

In this section, we give a simple unifying view of the most
popular graph data models, from the simplest ones used in graph databases to the models that have emerged to store, integrate and produce knowledge.

The following are two basic ingredients to define graph data models.
Assume that $\const$ is
a set of constants, or strings, that can be used for different
purposes, for example as node identifiers, edge identifiers, labels,
property names or actual values (such as integers, real values or dates). Moreover, define a multigraph 
as a 
graph where multiple edges can connect two
nodes, that is, a tuple $(N,E,\rho)$ where $N \subseteq \const$ is a
set of nodes, $E \subseteq \const$ is a set of edges and $\rho : E \to
N \times N$ is a function indicating the starting and ending node of
each edge. 

As a first data model, we consider labeled graphs, which are a popular
and simple way to represent semi-structured data. Formally, a labeled
graph is a tuple $\cL = (N,E,\rho,\lambda)$ where $(N,E,\rho)$ is a
multigraph and $\lambda : (N \cup E) \to \const$ is a function
indicating the label of each node and edge. 
Such graphs have been called heterogeneous graphs
in
the literature 
\cite{SHYYW11,DBLP:journals/corr/abs-2003-02320}, as opposed to
edge-labeled graphs where labels are only associated to
edges~\cite{AABHRV17}. But here we prefer the simple term labeled
graph to indicate that both nodes and edges are labeled. An example of
such a graph storing information about people and their contacts is
shown in Figure~\ref{fig-all-g}(a).











It is worth saying a few words about
RDF~\cite{world2014rdf}, a class of labeled graphs that is widely used
in practice. A first characteristic that distinguish RDF graphs is
that edges are replaced by triples, and they are not assigned
identifiers. Formally, an RDF graph is a set of triples $(s,p,o)$ such
that $s, p, o \in \const$, so that $(s,p,o)$ represents an edge from
$s$ to $o$ with label $p$. A second important feature of RDF graphs is
that $\const$ is considered as a set of Uniform Resource Identifiers
(URIs~\cite{berners1998uniform,durst2005internationalized}), that can
be used to identify any resource used by Web technologies. In this
way, RDF graphs have a universal interpretation: if
$c \in \const$ is used in two different RDF graphs, then $c$ is
considered to represent the same element.

As a second model we consider property graphs, which are widely
used in graph databases
\cite{miller2013graph,robinson2015graph,RHKMC16,FGGLLMPRS18}. Property graphs are defined as the extension of labeled graphs where
nodes and edges can have values for some properties. Formally, a
property graph is a tuple $\cP = (N,E,\rho,\lambda,\sigma)$ where
$(N,E,\rho,\lambda)$ is a labeled graph, and $\sigma : (N \cup E)
\times \const \to \const$ is a partial function such that if
$\sigma(o,p) = v$, then $v$ is said to be the value of property $p$
for object~$o$. Besides, it is assumed that 
each node or edge in $\cP$ has
values for a finite number of properties
\cite{miller2013graph,AABHRV17,AABBFGLPPS18}. In Figure~\ref{fig-all-g}(b),
we show an example of a property graph that extends the labeled graph in
Figure~\ref{fig-all-g}(a),
including as properties the name
and age of a person, the zip code of the address for two people
that live together, the date when someone rides a bus, and the date a contact between two people occurs.

As a final model, vector-labeled graphs are defined in a way that
unifies the use of labels and properties, and allows to include in a
simple way extra values that are necessary for message-passing graph
algorithms \cite{Kung82}, such as the Weisfeiler-Lehman
graph isomorphism test \cite{weisfeiler1968reduction,G11,GS20}, and
when graphs are used as input of graph neural networks
\cite{ML05,SGTHM09}. Formally, a vector-labeled graph of dimension $d$, with $d \geq 1$, is a tuple $\cV
= (N,E,\rho,\lambda)$ where $(N, E, \rho)$ is a multigraph and
$\lambda : (N \cup E) \to \const^d$ is a function that assigns a
vector of values to each node and edge
in the graph~\cite{DBLP:journals/corr/abs-2003-02320}, which is called a vector of features of dimension~$d$. Hence, labels and
properties are replaced by vectors of values from $\const$ in
vector-labeled graphs, as shown in Figure~\ref{fig-all-g}(c). In this figure, the string $\bot$ is used to represent the fact that a row in a vector does not have a value.

\section{Querying: some new challenges and techniques}
\label{sec:query}




The goal of this section is to present 
pointers to some recently obtained results on querying graphs, and to pose some research questions related to them, that could be of interest not only for the database community, but also for other communities like artificial intelligence and computational logic. In this sense, this will be a growing section divided according to the features identified in Section \ref{sec:what} for graph query languages. In particular, we consider the following problems in this version:
\begin{itemize}
    \item Local properties: we consider the old problem of extracting nodes from a graph, and present a recently established connection between graph neural networks and first-order logic with bounded resources, when such formalisms are viewed as query languages for node extraction.

\item Connectivity properties: we consider the old problem of extracting paths from a graph, and show some recent approximation and uniform generation results. 

\item Global properties: we provide a simple application of the approximation result mentioned in the previous point to the definition and computation of centrality measures. Moreover, we consider the area of Explainable AI~\cite{LL17,AB18,RSG18} and, in particular, the problem of defining a declarative language for expressing natural model interpretation tasks. More precisely, we argue that such a problem corresponds to the definition of a query language for some global properties in a graph database.
\end{itemize}


\subsection{A necessary terminology}
As mentioned before,
extracting nodes and paths is a fundamental task
when retrieving knowledge from graphs~\cite{AABHRV17,FGGLLMPRS18}. 
Regular expressions form the core of such an extraction task, so we need to fix a notation for regular expressions before going into the details of the results shown in this section.
More precisely, a regular expression over a labeled graph $\cL = (N,E,\rho,\lambda)$ is given by the following~grammar:
\begin{align}
\notag
    \test \ \ ::= &  \ \ \ \ell \ \mid\ (\neg \test) \ \mid\ (\test \vee \test) \ \mid\ (\test \wedge \test) \\
\label{eq-gramm}
    r \ \  ::= & \ \ \ ?\test  \ \mid\ \test \ \mid \ \test^-\ \mid\  (r+r) \ \mid\  (r/r) \ \mid\  (r^*),
\end{align}
where $\ell$ is a node or edge label in $\cL$. An answer to $r$ over $\cL$ is a path whose labels conform to $r$. Formally, such a path is a sequence $p = n_0 e_1 n_1 e_2 \cdots e_i n_i$, where $n_0, n_1, \ldots, n_i \in N$ and $e_1, \ldots, e_i \in E$. Moreover, the starting and ending nodes of $p$ are defined as $\stp(p) = n_0$ and $\ep(p) = n_i$, respectively, and the concatenation of $p$ with a path $p' = n_i e_{i+1} n_{i+1} e_{i+2} \cdots e_{i+j} n_{i+j}$ is defined as $\cat(p,p') = n_0 e_1 n_1 e_2 \cdots e_i n_i e_{i+1} n_{i+1} e_{i+2} \cdots e_{i+j} n_{i+j}$. With this terminology, the evaluation of $r$ over $\cL$, denoted by $\sem{r}$, is recursively defined as follows (omitting the usual interpretation for Boolean connectives $\neg$, $\vee$ and $\wedge$):
\begin{eqnarray*}
\sem{?\ell} & = & \{ n \mid n \in N \wedge \lambda(n) = \ell \}\\
\sem{\ell} & = & \{ n_0 e_1 n_1 \mid \rho(e_1) = (n_0,n_1) \wedge \lambda(e_1) = \ell \}\\
\sem{\ell^-} & = & \{ n_0 e_1 n_1 \mid \rho(e_1) = (n_1,n_0) \wedge \lambda(e_1) = \ell \}\\
\sem{r_1+r_2} & = & \sem{r_1} \cup \sem{r_2}\\
\sem{r_1/r_2} & = & \{ \cat(p_1,p_2) \mid p_1 \in \sem{r_1} \wedge p_2 \in \sem{r_2} \wedge\\
&&\phantom{\{ \cat(p_1,p_2) \mid\,}\ep(p_1) = \stp(p_2)\}\\
\sem{r^*} &=& N \cup \sem{r} \cup \sem{r/r} \cup \sem{r/r/r} \cup \cdots
\end{eqnarray*}
Notice that $?\ell$ is used to test the label of a node,  $\ell$ is used to follow an edge with label $\ell$, and $\ell^-$ is used to follow the opposite direction of an edge with label $\ell$. 
Besides, 
as an example of a test with Boolean connectives, observe that $\sem{(\neg \ell_1 \wedge \neg \ell_2)^-}
= \{ n_0 e_1 n_1 \mid \rho(e_1) = (n_1,n_0) \wedge
\lambda(e_1) \neq \ell_1 \wedge \lambda(e_1) \neq \ell_2\}$. 
Hence, if $\cL$ is the labeled graph in Figure \ref{fig-all-g}(a), then 
\begin{align}
\label{eq-re-1}
\sem{\lb{?person/contact/?infected}} =& \ \{ n_1 e_4 n_5 \},\\
\notag
\sem{\lb{?person/rides/?bus/}\lb{rides}^{\lb{-}}\lb{/?person}} =& \ \{n_1e_2n_3e_3n_4,\\
\notag
&\ \phantom{\{} n_4e_3n_3e_2n_1,\\
\notag
&\ \phantom{\{} n_1e_2n_3e_2n_1,\\
\notag
&\ \phantom{\{} n_4e_3n_3e_3n_4 \}.
\end{align}
As property graphs are an extension of labeled graphs, the grammar in \eqref{eq-gramm} can be easily expanded to consider property values:
\begin{align*}
\notag
    \test \ \ ::= &  \ \ \ \ell \ \mid \ (p=v) \ \mid \ (\neg \test) \ \mid \ 
    (\test \vee \test) \ \mid \ (\test \wedge \test).
\end{align*}
In particular, $(p=v)$ is used to verify whether the value of property $p$ is $v$, with $p,v \in \const$. Formally, the evaluation of a regular expression $r$ over a property graph $\cP = (N,E,\rho,\lambda,\sigma)$, denoted by $\semp{r}{\cP}$, is defined as for the case of labeled graphs but with three additional cases:
\begin{eqnarray*}
\semp{?(p = v)}{\cP} & = & \{ n \mid n \in N \wedge \sigma(n,p) = v \}\\
\semp{(p = v)}{\cP} & = & \{ n_0 e_1 n_1 \mid \rho(e_1) = (n_0,n_1) \wedge \sigma(e_1,p) = v \}\\
\semp{(p=v)^-}{\cP} & = & \{ n_0 e_1 n_1 \mid \rho(e_1) = (n_1,n_0) \wedge \sigma(e_1,p) = v \}.
\end{eqnarray*}
For example, we can extend regular expression \eqref{eq-re-1} to indicate that the date of the contact between a person and an infected person is March 4th 2021:
\begin{align}
\label{eq-re-pg}
\lb{?person/(contact}\wedge\lb{(date\,=\,3/4/21))/?infected}
\end{align}
Regular expressions for vector-labeled graphs are defined exactly in the same way. If $\cV =(N,E,\rho,\lambda)$ is a vector-labeled graph of dimension $d$, then a regular expression over $\cV$ is defined by modifying grammar \eqref{eq-gramm} to consider the following tests:
\begin{align*}
\notag
    \test \ \ ::= &  \ \ \ (f_i=v) \ \mid \ (\neg \test) \ \mid \ 
    (\test \vee \test) \ \mid \ (\test \wedge \test),
\end{align*}
where $i \in \{1, \ldots, d\}$ and $v \in \const$. 
In particular, $(f_i=v)$ is used to verify whether the value of the $i$-th feature is $v$, which is formally defined as follows:
\begin{eqnarray*}
\semp{?(f_i = v)}{\cV} & = & \{ n \mid n \in N \wedge \lambda(n)_i = v\}\\
\semp{(f_i = v)}{\cV} & = & \{ n_0 e_1 n_1 \mid \rho(e_1) = (n_0,n_1) \wedge \lambda(e_1)_i = v \}\\
\semp{(f_i=v)^-}{\cV} & = & \{ n_0 e_1 n_1 \mid \rho(e_1) = (n_1,n_0) \wedge \lambda(e_1)_i = v \},
\end{eqnarray*}
where $\lambda(n)_i$ refers to the $i$-th feature of $d$-dimensional vector $\lambda(n)$, and likewise for $\lambda(e_1)_i$. Thus, for example, regular expression \eqref{eq-re-pg} can be rewritten as follows over the vector-labeled graph in Figure~\ref{fig-all-g}(c):
\begin{align*}
\lb{?(f}_{\lb{1}}\lb{\,=\,person)/(f}_{\lb{1}}\lb{\,=\,contact}\wedge\lb{f}_{\lb{5}}\lb{\,=\,3/4/21)/?(f}_{\lb{1}}\lb{\,=\,infected)}.
\end{align*}

\subsection{Local properties}
\label{sec-p-m}

The task of matching a pattern against a graph is fundamental when extracting knowledge.
In the previous section, we 
considered this problem for regular expressions,
but such patterns can be specified in other frameworks, ranging from logic-based  declarative languages~\cite{HS13,B13} to more procedural frameworks such as graph neural networks \cite{ML05,SGTHM09}. The goal of this part of the document is to show a recently established tight connection between these apparently different frameworks~\cite{MRFHLRG19,XHLJ19,BKMPRS20}, which has interesting corollaries in terms of the use of declarative formalisms to specify patterns, versus the use of procedural formalisms to efficiently evaluate them.

Let $r_1 = \lb{?person/rides/?bus/}\lb{rides}^{\lb{-}}\lb{/?infected}$. How should this regular expression be evaluated over the labeled graph $\cL$ in Figure~\ref{fig-all-g}(a)? To think about this problem, let us focus on the task of retrieving the nodes $a$ from which a node $b$ can be reached by following a path conforming to~$r_1$, that is, a path $p \in \sem{r_1}$ such that $\stp(p) = a$ and $\ep(p) = b$. Pattern $r_1$ is then used to retrieve the list of people who are possibly infected because they shared a bus with infected people. This regular expression can be specified in first-order logic:
\begin{multline*}
\varphi(x) \ = \ 
    \lb{person}(x) \wedge \exists y \exists z \, (\lb{rides}(x,y) \wedge \lb{bus}(y) \ \wedge\\ \lb{rides}(z,y) \wedge \lb{infected}(z)),
\end{multline*}
considering node labels as unary predicates and edge labels as binary predicates. 
Notice that 
only unary and binary predicates are used in it, and they are placed in a sequence in which values of variables can be forgotten, allowing them 
to be reused. 
Indeed, the following first-order logic formula that uses two variables is equivalent to~$\varphi(x)$:
\begin{multline}
\label{eq-exa-tv}
\psi(x) \ = \ 
    \lb{person}(x) \wedge \exists y\, (\lb{rides}(x,y) \wedge \lb{bus}(y)
    \ \wedge \\ \exists x \, (\lb{rides}(x,y) \wedge \lb{infected}(x))).
\end{multline}
Thus, regular expression $r_1$ can be evaluated efficiently by noticing that only the values of variables $x$ and $y$ need to be stored when computing the answer to $\psi(x)$. Hence, the result of each join in $r$ is a binary table, and no auxiliary relations with an arbitrary number of columns have to be stored. 

The previous construction can be generalized to each regular expression $r$ not mentioning the Kleene star operator $(\ )^*$. That is, for each such an expression $r$, it is possible to construct a formula $\psi_r(x)$
in first-order logic such that $\psi_r(x)$ uses only two variables and for each labeled graph $\cL$ and node $a$ in $\cL$, there exists a path $p \in \sem{r}$ such that $\stp(p) = a$ if and only if $\psi_r(a)$ holds in $G$. We denote by $\FO^2$ the fragment of first-order logic where only two variables are allowed. Each formula in $\FO^2$ can be evaluated efficiently by generalizing the idea in the previous paragraph.
In fact, this idea has been successfully used in a variety of scenarios ~\cite{V95,AI00,GKP02,GKP05,M05},
and we are convinced that it 
should be kept in mind, not only as it provides an efficient way to evaluate regular expressions (without the Kleene star), but also as it allows to establish a tight connection with the more procedural and popular formalism of graph neural networks. 

\begin{figure*}

\begin{center}
\resizebox{\textwidth}{!}{
\begin{tikzpicture}
\node[rt] (vn1) {$n_1$};

\node[rectw, above left=0mm of vn1] (vvn1) {
{\scriptsize
$\begin{pmatrix}
  \lb{bus}\\
  \lb{0}
\end{pmatrix}$}};

\node[rt, above right=1cm and 2.5cm of vn1] (vn2) {$n_2$};

\node[rectw, above right=0mm of vn2] (vvn2) {
{\scriptsize
$\begin{pmatrix}
  \lb{infected}\\
  \lb{0}
\end{pmatrix}$}};

\draw[arrin,bend left=10] (vn1) to 
node[ert, above] (ve1) {$e_1$}
(vn2);

\node[rectw, above=0mm of ve1] (vve1) {
{\scriptsize
$\begin{pmatrix}
  \lb{rides}\\
  \lb{0}
\end{pmatrix}$}};

\node[rt, below=2cm of vn1] (vn3) {$n_3$};

\node[rectw, below left=0mm of vn3] (vvn3) {
{\scriptsize
$\begin{pmatrix}
  \lb{person}\\
  \lb{0}
\end{pmatrix}$}};

\draw[arrin,bend right=10] (vn1) to 
node[ert, right] (ve2) {$e_2$}
(vn3);

\node[rectw, right=0mm of ve2] (vve2) {
{\scriptsize
$\begin{pmatrix}
  \lb{rides}\\
  \lb{0}
\end{pmatrix}$}};

\node[rt, right=3.5cm of vn3] (vn4) {$n_4$};

\node[rectw, below right=0mm of vn4] (vvn4) {
{\scriptsize
$\begin{pmatrix}
  \lb{bus}\\
  \lb{0}
\end{pmatrix}$}};

\node[rt, above=2cm of vn4] (vn5) {$n_5$};

\node[rectw, right=0mm of vn5] (vvn5) {
{\scriptsize
$\begin{pmatrix}
  \lb{person}\\
  \lb{0}
\end{pmatrix}$}};

\draw[arrin, bend left=10] (vn4) to 
node[ert, below] (ve3) {$e_3$}
(vn3);

\node[rectw, below=0mm of ve3] (vve3) {
{\scriptsize
$\begin{pmatrix}
  \lb{rides}\\
  \lb{0}
\end{pmatrix}$}};

\draw[arrout, bend left=10] (vn5) to 
node[ert, left] (ve4) {$e_4$}
(vn4);

\node[rectw, left=0mm of ve4] (vve4) {
{\scriptsize
$\begin{pmatrix}
  \lb{rides}\\
  \lb{0}
\end{pmatrix}$}};

\node[rtr, right=32mm of vn5] (svn1) {{\color{red} $n_1$}};

\node[rectw, above left=0mm of svn1] (svvn1) {
{\color{red}
{\scriptsize
$\begin{pmatrix}
  \lb{bus}\\
  \lb{1}
\end{pmatrix}$}}};

\node[rt, above right=1cm and 2.5cm of svn1] (svn2) {$n_2$};

\node[rectw, above right=0mm of svn2] (svvn2) {
{\scriptsize
$\begin{pmatrix}
  \lb{infected}\\
  \lb{0}
\end{pmatrix}$}};

\draw[arrin,bend left=10] (svn1) to 
node[ert, above] (sve1) {$e_1$}
(svn2);

\node[rectw, above=0mm of sve1] (svve1) {
{\scriptsize
$\begin{pmatrix}
  \lb{rides}\\
  \lb{0}
\end{pmatrix}$}};

\node[rt, below=2cm of svn1] (svn3) {$n_3$};

\node[rectw, below left=0mm of svn3] (svvn3) {
{\scriptsize
$\begin{pmatrix}
  \lb{person}\\
  \lb{0}
\end{pmatrix}$}};

\draw[arrin,bend right=10] (svn1) to 
node[ert, right] (sve2) {$e_2$}
(svn3);

\node[rectw, right=0mm of sve2] (svve2) {
{\scriptsize
$\begin{pmatrix}
  \lb{rides}\\
  \lb{0}
\end{pmatrix}$}};

\node[rt, right=3.5cm of svn3] (svn4) {$n_4$};

\node[rectw, below right=0mm of svn4] (svvn4) {
{\scriptsize
$\begin{pmatrix}
  \lb{bus}\\
  \lb{0}
\end{pmatrix}$}};

\node[rt, above=2cm of svn4] (svn5) {$n_5$};

\node[rectw, right=0mm of svn5] (svvn5) {
{\scriptsize
$\begin{pmatrix}
  \lb{person}\\
  \lb{0}
\end{pmatrix}$}};

\draw[arrin, bend left=10] (svn4) to 
node[ert, below] (sve3) {$e_3$}
(svn3);

\node[rectw, below=0mm of sve3] (svve3) {
{\scriptsize
$\begin{pmatrix}
  \lb{rides}\\
  \lb{0}
\end{pmatrix}$}};

\draw[arrout, bend left=17] (svn5) to 
node[ert, left] (sve4) {$e_4$}
(svn4);

\node[rectw, left=0mm of sve4] (svve4) {
{\scriptsize
$\begin{pmatrix}
  \lb{rides}\\
  \lb{0}
\end{pmatrix}$}};

\node[rt, right=32mm of svn5] (tvn1) {$n_1$};

\node[rectw, above left=0mm of tvn1] (tvvn1) {
{\scriptsize
$\begin{pmatrix}
  \lb{bus}\\
  \lb{1}
\end{pmatrix}$}};

\node[rt, above right=1cm and 2.5cm of tvn1] (tvn2) {$n_2$};

\node[rectw, above right=0mm of tvn2] (tvvn2) {
{\scriptsize
$\begin{pmatrix}
  \lb{infected}\\
  \lb{0}
\end{pmatrix}$}};

\draw[arrin,bend left=10] (tvn1) to 
node[ert, above] (tve1) {$e_1$}
(tvn2);

\node[rectw, above=0mm of tve1] (tvve1) {
{\scriptsize
$\begin{pmatrix}
  \lb{rides}\\
  \lb{0}
\end{pmatrix}$}};

\node[rtr, below=2cm of tvn1] (tvn3) {{\color{red} $n_3$}};

\node[rectw, below left=0mm of tvn3] (tvvn3) {
{\scriptsize
{\color{red}
$\begin{pmatrix}
  \lb{person}\\
  \lb{1}
\end{pmatrix}$}}};

\draw[arrin,bend right=10] (tvn1) to 
node[ert, right] (tve2) {$e_2$}
(tvn3);

\node[rectw, right=0mm of tve2] (tvve2) {
{\scriptsize
$\begin{pmatrix}
  \lb{rides}\\
  \lb{0}
\end{pmatrix}$}};

\node[rt, right=3.5cm of tvn3] (tvn4) {$n_4$};

\node[rectw, below right=0mm of tvn4] (tvvn4) {
{\scriptsize
$\begin{pmatrix}
  \lb{bus}\\
  \lb{0}
\end{pmatrix}$}};

\node[rt, above=2cm of tvn4] (tvn5) {$n_5$};

\node[rectw, right=0mm of tvn5] (tvvn5) {
{\scriptsize
$\begin{pmatrix}
  \lb{person}\\
  \lb{0}
\end{pmatrix}$}};

\draw[arrin, bend left=10] (tvn4) to 
node[ert, below] (tve3) {$e_3$}
(tvn3);

\node[rectw, below=0mm of tve3] (tvve3) {
{\scriptsize
$\begin{pmatrix}
  \lb{rides}\\
  \lb{0}
\end{pmatrix}$}};

\draw[arrout, bend left=17] (tvn5) to 
node[ert, left] (tve4) {$e_4$}
(tvn4);

\node[rectw, left=0mm of tve4] (tvve4) {
{\scriptsize
$\begin{pmatrix}
  \lb{rides}\\
  \lb{0}
\end{pmatrix}$}};

\node[rectw, below=1.5cm of ve3] (ml0) {(a) Layer 0};

\node[rectw, below=1.5cm of sve3] (ml1) {(b) Layer 1};

\node[rectw, below=1.5cm of tve3] (ml2) {(b) Layer 2};

\end{tikzpicture}
}
\end{center}

\caption{Execution of a graph neural network. \label{fig-exe-gnn}}
\end{figure*}
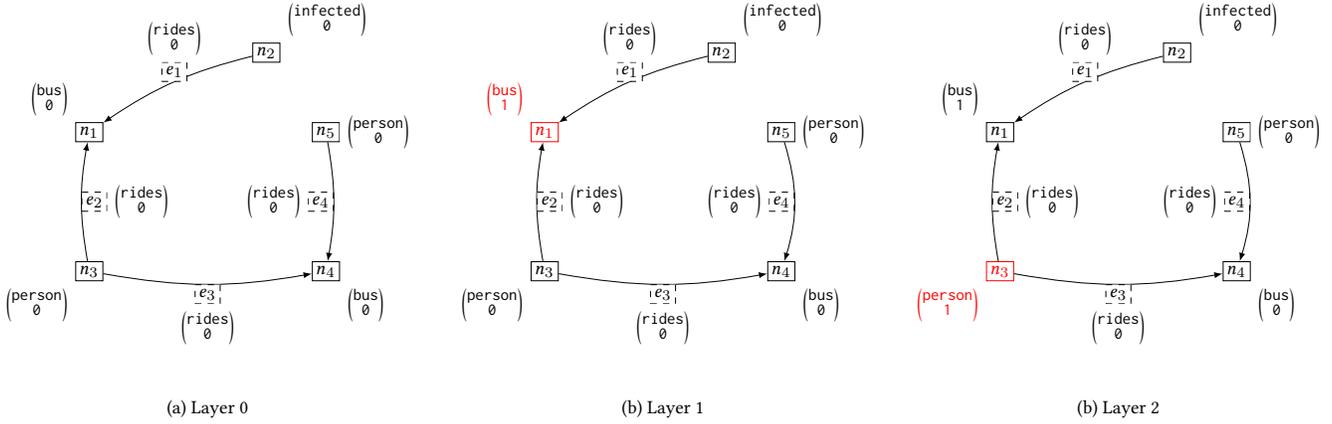

A graph neural network (GNN) receives as input a vector-labeled graph $\cV = (N,E,\rho,\lambda)$ of dimension $d$, generates from it a vector-labeled graph $\cV' = (N,E,\rho,\lambda')$ of the same dimension $d$, and then
uses 
a classification function that returns 
either 
\textit{true} or \textit{false} for each node 
$u \in N$ based on 
the vector 
$\lambda'(u)$.
Formally, a GNN $\cG$ with $\ell \geq 1$ layers is given by a sequence of functions $\AGG^{(i)}$ and $\COMB^{(i)}$ for $i \in \{1, \ldots, \ell\}$, and a function $\CSL$. For each layer $i \in \{0, 1, \ldots, \ell\}$, a vector $u^{(i)} \in \const^d$ is associated to each node $u \in N$. In particular, the input graph $\cV = (N,E,\rho,\lambda)$ is considered as the layer $0$ for $\cG$, so that $u^{(0)} = \lambda(u)$. Moreover, assuming that layer $i$ has been computed, the layer $i+1$ is defined as follows. A node $v$ is said to be a neighbor of a node $u$ in $\cV$ if there exists $e \in E$ such that $\rho(e) = (u,v)$ or $\rho(e) = (v,u)$. Then 
\begin{multline*}
u^{(i+1)}  \ =\  \COMB^{(i)}\big(u^{(i)},\\
\AGG^{(i)}(\{\!\!\{ v^{(i)} \mid v \text{ is a neighbor of } u \text{ in } \cV\}\!\!\})\big),
\end{multline*}
where $\AGG^{(i)}$ is an aggregate function that takes as input the multiset of feature vectors of the neighbors of node $u$ at layer $i$ and produces a vector $r^{(i)} \in \const^d$, and $\COMB^{(i)}$ is a function that combines the feature vector of node $u$ at layer $i$ with $r^{(i)}$ to produce the feature vector of node $u$ at layer $i+1$. Finally, the output of $\cG$ for a node $u$, denoted by $\cG(u,\cV)$,  is defined as $\CSL(u^{(\ell)})$, that is, as the result of applying the classification function $\CSL$ over the last layer computed by the graph neural~network. 

For example, consider the vector-labeled graph $\cV$ in Figure~\ref{fig-exe-gnn}(a), and a GNN $\cG$ with two layers used to identify the list of people from $\cV$ who are possibly infected because they shared a bus with infected people. The layer 0 of the computation is shown in Figure~\ref{fig-exe-gnn}(a), which corresponds to the input graph $\cV$. Then $\cG$ computes $u^{(1)}$ as follows, for each node $u$ of $\cV$. If the first component of $u^{(0)}$ is \lb{bus} and $u$ has a neighbor $v$ such that the first component of $v^{(0)}$ is \lb{infected}, then $u^{(1)}$ is defined as $u^{(0)}$ but with the second component changed to $1$. Otherwise, $u^{(1)} = u^{(0)}$. In this way, if an infected person rode a bus, then this bus is signaled in layer 1, as shown by the node marked in red in Figure \ref{fig-exe-gnn}(b). Notice that the previous computation is obtained by combining functions $\AGG^{(1)}$ and $\COMB^{(1)}$, which are not explicitly defined for the sake of presentation.
Moreover, 
$\cG$ computes $u^{(2)}$ as follows, for each node $u$ of $\cV$. If the first component of $u^{(1)}$ is \lb{person} and $u$ has a neighbor $v$ such that the first component of $v^{(1)}$ is \lb{bus} and the second component of $v^{(1)}$ is $1$, then $u^{(2)}$ is defined as $u^{(1)}$ but with the second component changed to $1$. Otherwise, $u^{(2)} = u^{(1)}$. In this way, possibly infected people (because they shared a bus with infected people) are identified in layer 2, as shown by the node marked in red in Figure \ref{fig-exe-gnn}(c). Finally, the classification function $\CSL$ assigns value \textit{true} to a node $u$ if, and only if, the first component of $u^{(2)}$ is \lb{person} and the second component of $u^{(2)}$ is $1$. Hence, we have that $\cG(n_3, \cV) = \textit{true}$, and 
$\cG(n_1, \cV) = \cG(n_2, \cV) = \cG(n_4, \cV) = \cG(n_5, \cV) = \textit{false}$.

Notice that the previous architecture for graph neural networks takes into account neither the direction nor the label of an edge. If the directions of the edges of a vector-labeled graph $\cV = (N,E,\rho,\lambda)$ have to be considered, then $\AGG^{(i)}$ has to receive as separate inputs the multisets:
\begin{align*}
&\{\!\!\{ v^{(i)} \mid \rho(e) = (u,v) \text{ for some edge } e \in E \}\!\!\},\\
&\{\!\!\{ v^{(i)} \mid \rho(e) = (v,u) \text{ for some edge } e \in E \}\!\!\}.
\end{align*}
Moreover, if the labels of edges have to be taken into account, then a vector of features $e^{(i)}$ has to be associated with each edge $e$, which is considered in the definition of a GNN in the same way as the vector of features $u^{(i)}$ for a node $u$.
 
The architecture for a graph neural network $\cG$ defined in this section, that is referred to as an aggregate-combine graph neural network~\cite{BKMPRS20}, turns $\cG$ into a classifier~\cite{ML05,SGTHM09}.
But also 
$\cG$ can 
be considered as a 
unary 
query
that 
is true for a node $u$ of a vector-labeled graph $\cV$ if, and only if, the output of $\cG$ is $\textit{true}$ for $u$, that is $\cG(u,\cV) = \textit{true}$. For example, the graph neural network depicted in Figure \ref{fig-exe-gnn} corresponds to the unary query defined by first-order formula $\psi(x)$ in \eqref{eq-exa-tv}. Hence, it is important to understand the expressiveness of graph neural networks as a query language, in particular because they can act as an efficient procedural counterpart of more declarative query formalisms. 

Define $\FOC^2$ as the extension of $\FO^2$ with quantifiers of the form $\exists x^{\geq k} x \, \psi(x)$ with $k$ a positive integer number, which is satisfied if $\varphi(a)$ holds for at least $k$ distinct values $a$ for variable $x$. Notice that $\FOC^2$ is an extension of $\FO^2$ as quantifier $\exists x^{\geq k} x \, \psi(x)$ can be expressed in first-order logic but using $k$ variables. 
In  \cite{BKMPRS20}, it is proved that there is a tight connection between a natural fragment of $\FOC^2$ and aggregate-combine GNNs, since these two formalisms have the same expressive power as query languages. Such a characterization is based on a classical result showing that the Weisfeiler-Lehman test for graph isomorphism \cite{weisfeiler1968reduction} has the same expressive power as $\FOC^2$~\cite{CFI92}, and a recently established result showing that the  Weisfeiler-Lehman test bounds the expressiveness of aggregate-combine GNNs~\cite{MRFHLRG19,XHLJ19} .
Interestingly, the Weisfeiler-Lehman
test can be formalized as 
a message-passing graph
algorithm \cite{Kung82}, which is an algorithmic model intimately related with graph neural networks.
For a more detailed, and formal, description of the results outlined in this section see \cite{BKMPRS20,DBLP:journals/corr/abs-2104-14624}, where also many other interesting questions about the relationship between formal logic and GNNs are posed.

We conclude this section by mentioning some research questions that may be of general interest. Can learning techniques for GNNs be used for learning queries on graphs? What are good algorithms for translating aggregate-combine GNNs into well-studied declarative query languages? And what is an appropriate GNN architecture for regular expressions? Notice that the answer to this last question will need of a query language with some form of recursion, which is a feature not expressible in $\FO$~\cite{AHV95,L04}.

\subsection{Connectivity properties}
\label{sec-ug-c}

Computing the complete set of answers to a graph query can be prohibitively expensive~\cite{ACP12,LM13}. 
As a way to overcome this limitation, the idea of enumerating the answers to a query with a small delay has recently attracted much attention 
\cite{S13,MT19}. More specifically, the computation of the answers is divided 
into 
a preprocessing phase, where a data structure is built to accelerate the process of computing answers, and then in an enumeration phase, the answers are produced with a polynomial-time
delay between them.


Unfortunately, because of the data structures used in the preprocessing phase, these enumeration algorithms usually return answers that are similar to each other \cite{bagan2007acyclic,S13,florenzano2018constant}.  
In this respect, the possibility of generating an answer uniformly, at random, is a desirable condition to improve the variety, if it can be done efficiently~\cite{AMSW16,DBLP:conf/nips/Har-PeledM19,AD20,DBLP:conf/pods/0001P020}.
However, 
how can we know how complete is the set of answers calculated by such algorithms?
A third tool that is needed then is an efficient algorithm for computing, or estimating, the number of solutions to a query. 

In the following, we will present some recent results on efficient enumeration, uniform generation and approximate counting of paths conforming to a regular expression~\cite{ACJR19,ACJR20}.
We will give an overview of two of these results for labeled graphs, but the reader must keep in mind that they can be readily adapted to  property and vector-labeled graphs.

The length of a path $p = n_0 e_1 n_1 e_2 \cdots e_k n_k$, denoted by $|p|$, is defined as~$k$. The problem $\ccp$ has as input a labeled graph $\cL$, a regular expression $r$ over $\cL$ and a number $k$ (given in unary as a string $0^k$), and the task is to compute the number of paths $p \in \sem{r}$ with $|p| = k$, which is denoted by $\ccp(G,r,k)$. The problem $\ccp$ is known to be intractable; in fact, it is  $\spanl$-complete~\cite{AJ93}, which implies that if $\ccp$ can be solved in polynomial time, then $\p = \np$~\cite{AJ93}. However, it has been recently shown that $\ccp$ can be efficiently approximated~\cite{ACJR19}. More precisely, it has been shown that there exists a randomized algorithm $\cA$ that receives as input $\cL$, $r$, $k$ and an error $\varepsilon \in (0,1)$, and computes a value $\cA(G,r,k,\varepsilon)$ such that:
\begin{eqnarray*}
\Pr\bigg(\,\bigg|\frac{\ccp(G,r,k) - \cA(G,r,k,\varepsilon)}{\ccp(G,r,k)}\bigg| \,\leq\, \varepsilon\,\bigg) & \geq & 1-\bigg(\frac{1}{2}\bigg)^{100},
\end{eqnarray*}
that is, with a very high probability the algorithm returns a value whose relative error is at most $\varepsilon$. Moreover, the algorithm works in polynomial time in the size of $\cL$, $r$, and the values $k$ and $\nicefrac{1}{\varepsilon}$. Notice that such an algorithm is a fully polynomial-time randomized approximation scheme for $\ccp$~\cite{DBLP:books/daglib/0004338}.

The problem $\gen$ has the same input $\cL$, $r$, $k$ as $\ccp$,
but the task is to generate uniformly, at random, a path $p \in \sem{r}$ with $|p| = k$. As a corollary of the previous result, it is obtained that $\gen$ can be solved efficiently.
More precisely, there exists a randomized algorithm $\cB$ that is divided into a preprocessing and a generation phase. In the preprocessing phase, the algorithm construct with a very high probability a data structure, 
which can be repeatedly used in the generation phase to produce paths $p \in \sem{r}$ of length $k$ with uniform distribution.
  
Some questions about the results outlined in this section need to be answered. In particular, whether the fully polynomial-time randomized approximation scheme for $\ccp$ can be effectively used in practice, and whether it can be used to provide {\em fair} answers to a query based on regular expressions.

\subsection{Global properties}
\label{sec-g-a}
Graph analytic makes reference to a series of techniques to analyze the structure and content of a graph as a whole. Typical applications include clustering~\cite{S07}, 
computation of connected components and the diameter of a graph, 
computation of shortest paths between pairs of nodes, 
calculation of centrality measures~\cite{newman2018networks}, such as betweenness centrality \cite{freeman1977set} and PageRank \cite{BP98}, 
and community detection, such as finding the subgraph of a graph with the largest density \cite{goldberg1984finding,MFCL0020}, to identify groups with a rich interaction in a network \cite{K99} or groups with suspicious  behaviour~\cite{PSSMF10,HSBSSF16}.

How should knowledge be included in such techniques? We focus here on the task of computing centrality. 
Given a labeled graph $\cL = (N,E,\rho,\lambda)$ and nodes $a, b, x \in N$, let $S_{a,b}$ be the set of shortest paths from $a$ to $b$ in $\cL$, and $S_{a,b}(x)$ be the set of paths in $S_{a,b}$ including node $x$. Then the betweenness centrality of a node $x$ of $\cL$ is defined as \cite{freeman1977set}:
\begin{eqnarray*}
\bc(x) &=& \sum_{a,b \in N \,:\, a\neq x \wedge b\neq x} \frac{|S_{a,b}(x)|}{|S_{a,b}|}
\end{eqnarray*}
This definition does not use the labels in $\cL$, which may be a problem if not all the shortest paths passing through a node need to be considered to measure its centrality. Of course, not including some nodes and edges in the computation can be a solution to this problem. But, unfortunately, in many cases this is not enough as the pattern defining the paths to be taken into account can be more complicated. As an example, consider the labeled graph in Figure~\ref{fig-all-g}(a), and assume that we want to measure the centrality of bus $n_3$ as a transportation service with respect to other buses. In this case, we should only consider the shortest paths confirming to the regular expression $r = \lb{?person/rides/?bus/}\lb{rides}^{\lb{-}}\lb{/?person}$.
That is, we must consider the paths where the bus is used as a transportation service for people, and not, for example, the paths with information about the company that owns it.
In fact, if $S_{a,b,r}$ is the set of shortest paths from $a$ to $b$ conforming to the regular expression $r$, and $S_{a,b,r}(n_3)$ is the set of paths in $S_{a,b,r}$ including node $n_3$, then the centrality of $n_3$ can be redefined as~follows:
\begin{eqnarray*}
\bc_r(n_3) &=& \sum_{a,b \in N \,:\, a\neq n_3 \wedge b\neq n_3} \frac{|S_{a,b,r}(x)|}{|S_{a,b,r}|}
\end{eqnarray*}
This definition can be generalized to any regular expression $r$.
For example, the 
regular expression $r_1 = \lb{?infected/rides/}\lb{?bus/}\lb{rides}^{\lb{-}}\lb{/}$
$\lb{(?person/}\lb{(lives}+\lb{contact))}^{\lb{*}}\lb{/?person}$
can be used in conjunction with betweenness centrality to measure the important of a bus in the propagation of an infection.
In fact, $r_1$ is used to find pairs $(a,b)$ of people such that $a$ is infected and shared a bus with a person $c$, and $b$ is connected to $c$ through a path of arbitrary length of people that lives together or have been in contact with each other.

Betweenness centrality can be computed efficiently, as there exists an 
efficiently algorithm for the following problem: given a labeled graph $\cL$, a pair of nodes $a,b$ in $\cL$ and a length $k$, count the number of paths of length $k$ from $a$ to $b$ in~$\cL$. However, as mentioned in Section \ref{sec-ug-c}, the situation is different if regular expressions are considered as the previous problem is intractable \cite{AJ93}.
How can we overcome this limitation? We leave as an exercise for the reader to show that the tools presented in Section \ref{sec-ug-c} 
can be used to provide an efficient randomized approximation algorithm for $\bc_r(\cdot)$.

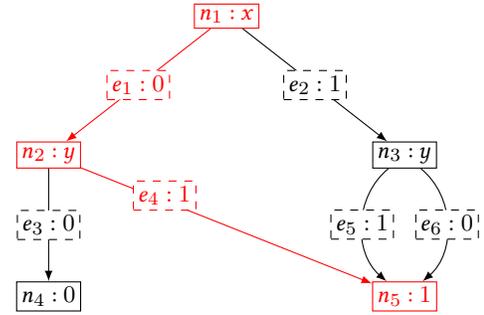
\begin{figure}[h]

\begin{center}
\begin{tikzpicture}
\node[rtr] (n1) {{\color{red} $n_1$ : $x$}};

\node[rtr, below left=1.5cm and 1.5cm of n1] (n2) {{\color{red} $n_2$ : $y$}};

\node[rt, below right=1.5cm and 1.5cm of n1] (n3) {$n_3$ : $y$};

\node[rt, below=1.5cm of n2] (n4) {$n_4$ : $0$};

\node[rtr, below=1.5cm of n3] (n5) {{\color{red} $n_5$ : $1$}};

\draw[arrout,color=red] (n1) to 
node[ertr] (e1) {{\color{red} $e_1$ $:$ $0$}}
(n2);

\draw[arrout] (n1) to 
node[ert] (e2) {$e_2$ $:$ $1$}
(n3);

\draw[arrout] (n2) to 
node[ert] (e3) {$e_3$ $:$ $0$}
(n4);

\draw[arrout,color=red] (n2) to 
node[ertr,above left=2mm and 4mm] (e4) {{\color{red} $e_4$ $:$ $1$}}
(n5);

\draw[arrout, bend right=50] (n3) to 
node[ert] (e5) {$e_5$ $:$ $1$}
(n5);

\draw[arrout, bend left=50] (n3) to 
node[ert] (e4) {$e_6$ $:$ $0$}
(n5);

\end{tikzpicture}
\end{center}

\caption{A decision model represented as a labeled graph. \label{fig-dm-lg}}
\end{figure}

As in the previous sections, we would like to conclude by pointing out some questions for future research.
In particular, it is a challenging question how knowledge should be considered in centrality measures.
In a recent article \cite{RS20}, the authors provide a natural and general framework to specify centrality measures, where betweenness centrality can be defined, but still without taking labels into consideration. Moreover, it is also important to mention a connection between the notion of global property in a graph database and the definition of a declarative language for expressing natural model interpretation tasks, in the area of Explainable AI~\cite{LL17,AB18,RSG18}. 

Consider the labeled graph $\cL$ in Figure~\ref{fig-dm-lg}, in which node labels are used to represent variables in a binary decision model $\cM$, and edge labels are used to represent assignments for such variables. For example, the path marked in red in Figure~\ref{fig-dm-lg} corresponds to an instance where $x$ is assigned value $0$ and $y$ is assigned value $1$. Besides, nodes $n_4$ and $n_5$ are used in $\cL$ to denote the value given by $\cM$ to an instance; for example, the instance represented by the path marked in red in $\cL$ is assigned value $1$ by the model. Several queries can be used when trying to understand the way in which $\cM$ classifies instances. The most simple ones ask whether there is any instance that is classified positively by $\cM$, whether there is any instance that is classified negatively by the model, or whether a partial instance like $x \mapsto 0$ can be completed to obtain an instance classified positively. Notice that such queries retrieve paths satisfying some conditions, so they correspond to what we have called connectivity properties in this document. 

However, other interpretation tasks need to consider global properties of the labeled graph representing a decision model. For example, in the decision model $\cM$ depicted in Figure~\ref{fig-dm-lg}, if variable $x$ is assigned value $1$ in an instance, then such an instance is classified positively by $\cM$, no matter what value is assigned to variable $y$. In this sense, $x \mapsto 1$ can be considered as a sufficient reason for a positive classification of an instance, as any completion of $x \mapsto 1$ is assigned value $1$ by the model. Following this idea, several relevant queries about the global structure of the graph may be posed. Given an instance classified positively, what is a sufficient reason for it? What is a minimal sufficient reason for this instance? What are the minimal sufficient reasons in the model? Is the model biased with respect to a protected feature?

As a final remark, it is an interesting open problem how to define a declarative language for model interpretability, which not only should be expressive enough to represent the aforementioned interpretation queries, 
but also should be adequate for efficient implementation. Notice that this latter problem is directly related with the representation used for classification models, and the structural properties of this representation. For instance, the labeled graph representing a classification model can be a decision tree~\cite{rudin2019stop}, in which case the aforementioned interpretation queries can be solved efficiently~\cite{BM0S20}.

\section{Takeaway messages}
\label{sec:takeaway}

The richness of the manifold technical developments in the area, part of which we reviewed,
 deserves to be encompassed in a conceptual map.
Graphs have become ubiquitous in data and knowledge management. 
    We argued that one of the main drivers of this blooming is 
the dual character of graphs: on one hand, being a simple, flexible and 
extensible data structure; and on the other, being one of the most
deep-rooted form of representing human knowledge. 


Graphs (as representation) unveil social aspects that are
relatively far from being the main concerns of our area. Traditionally, we
divided our labor between designers, organizing the conceptual boxes
(through schemata and metadata) that contain data; and we, data people,
dealing with preserving and transforming such data. 
Knowledge graphs mixed both worlds making difficult to trace a clear frontier between them.

Today we witness highly efficient techniques, particularly 
from the area of statistics, that are coping our area. They have to do with
the massive and automated collection of new type of data, 
thus unstructured and uncertain.
We have been using them 
mainly as tools, but large graphs 
force to incorporating them harmonically into our discipline.
Much of this has to do with the classic counterpoint between logic and statistics,
but it goes much deeper.

Finally, we are convinced that we should re-think the very notion of ``querying'' in graphs. We try to organize current
research in three big areas, namely entities/nodes, relationships/connectivity,
and emergent/global properties. But orthogonal to this 
is the extension of classical queries as 
languages for transforming data into
data plus a final interpretation, into a loop with a continuous process of interaction between humans and data.
Probably the allure of graphs today has to do with this loop.


\begin{acks}
This work was funded by ANID - Millennium Science Initiative Program - Code ICN17\_002.
\end{acks}

\balance

\bibliographystyle{ACM-Reference-Format}
\bibliography{biblio}

\end{document}